\documentclass[10pt]{article}
\usepackage{sao2}
\usepackage{psfig,epsf}

\issue{2009, 64, 357--365}

\begin{document}
\markboth{M.L.~Khabibullina, O.V.~Verkhodanov}
     {CATALOG OF RADIO GALAXIES WITH $z>0.3$. III: Angular Sizes and Flux Density According to the NVSS Data}
\title{Catalog of Radio Galaxies with~ $z>0.3$. III: Angular Sizes and Flux Density According to the NVSS Data}
\author{M.L.~Khabibullina\inst{a}
\and O.V.~Verkhodanov\inst{a}
}
\institute{
\saoname
}
\date{December 15, 2008}{April 4, 2009}
\maketitle

\begin{abstract}
We describe the procedure of the construction of a sample of
distant ($z>0.3$) radio galaxies  using the NED, CATS and SDSS
databases for further use in various statistical tests. We believe
the sample to be free of objects with quasar properties. This
paper is the third part of the description of the radio galaxies
catalog that we plan to use for cosmological tests. We report the
results of the sample of angular sizes for the NVSS survey list
objects, and its preliminary statistical analysis. Three-parameter
diagrams ``angular size--redshift--flux density'' and ``angular
size--redshift--spectral index'', and their two-parameter
projections are constructed. Three subsamples of radio galaxies
are separated in the ``source size--flux density'' diagram.
\mbox{\hspace{1cm}}
%\pacs{98.54.Gr, 98.62.Py, 98.70.Dk, 98.80.Es} \maketitle
\end{abstract}

\section{INTRODUCTION}

The present work is the third part of the catalog of radio
galaxies located within $0.3<z<5.2$ redshifts. The first two parts
\cite{rg_listI:Verkhodanov3_n_en,rg_listII:Verkhodanov3_n_en} (further, Papers I and II) describe the
way  the sample was constructed, based on the NED\footnote{\tt
http://nedwww.ipac.caltech.edu}, CATS\footnote{\tt
http://cats.sao.ru}, and SDSS\footnote{\tt http://sdss.org} data,
including an analysis of other works with the aim of excluding the
objects with quasar properties. Paper~I lists 2442 objects with
their equatorial coordinates for epoch J2000.0, spectroscopic
redshifts, and spectral indices, deduced according to the
cross-identification results of the selected radio galaxies with
the CATS radio catalogs, and computed at 325, 1400 and 4850\,MHz.
The list of objects in Paper~I also contains the names of the
initial catalogs from where we adopted the data for the objects
(and the names of radio sources in case of the 3C and 4C surveys).
Paper~II presents photometric data from the NED and SDSS archives.
We constructed the ``magnitude--redshift'' Hubble diagrams for 11
filters. For the SDSS survey \cite{sdss:Verkhodanov3_n_en} (g, i, r, z passbands)
we determined the parameters of linear regression on the Hubble
diagram. The preparation of the catalog presented in the three
papers is the first step for making cosmological tests on the
observed physical parameters of radio galaxies at different
epochs.

An  objective of our catalog was a manifold increase of the
comparatively uniform sample of radio galaxies, which would allow
increasing the accuracy of the determination of cosmological
parameters in an independent set of tests, apart from the
cosmic microwave background and a large-scale structure. In the
last part of the catalog, which is supplemented with this paper,
we give the list of objects with their angular sizes from the NVSS
survey\footnote{\tt http://www.cv.nrao.edu/nvss/} \cite{nvss:Verkhodanov3_n_en} and
the flux density measurements data at 1400\,MHz.

The radio source characteristics collected in this list can be
used in the ``radio luminosity--redshift'' and ``angular
size--redshift'' tests. Both tests are connected with the galaxy
characteristics that greatly evolve. Hence, the evaluation of the
cosmological parameters, based on them, has a large dispersion.
Nevertheless, the ``angular size--redshift'' (or
\mbox{``$\theta-z$'')} diagram is widely used in the papers, that
consider both the minimal size of point sources \cite{gurvits:Verkhodanov3_n_en},
and the average angular size of a radio galaxy at a given
redshift, corrected for evolutionary effects \cite{guerra:Verkhodanov3_n_en}. The
angular size of a radio source depending on its redshift can be
used in a combined test with the spot sizes in the cosmic
microwave background ~\cite{jackson_janneta:Verkhodanov3_n_en}, where both the
scale of the CMB fluctuations, and the characteristic sizes of
millisecond sources are used.

Note once more that all the methods of parameter finding from the
relation ``$\theta-z$'' are aggravated with uncertainties linked
with evolutionary effects, and the

\begin{figure*}[tbp]
\centerline{\hbox{
\psfig{figure=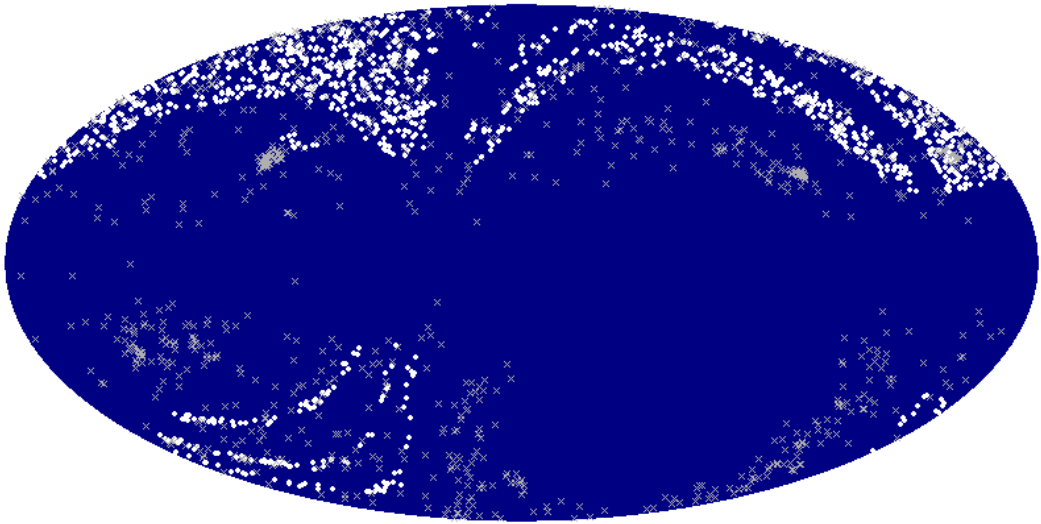,width=14cm}
}}
\caption{Sky positions of selected radio sources in Galactic
coordinates. Objects location is determined by the simultaneous
presence of observational data in optical and radio ranges. White
circles indicate SDSS objects and gray crosses mark all other
sources.}
\label{f1:Verkhodanov3_n_en}
\end{figure*}

\noindent extended objects, in addition, with the task of
determining the outer edge of an object. Nevertheless, using the
source size as a standard ruler is justified. If radio galaxies
have a fundamental plane, determined by such physical parameters
as, e.g., the mass of the radio galaxy or its central black hole,
which was already mentioned in Paper\,II, then we have reason to
link the angular size of the source with the history of formation
of the active nucleus, described by a set of standard measurable
parameters, like the accretion velocity, angular momentum,
magnetic field magnitude, and mass of the central black hole. This
gives us grounds to expect that the proposed in the literature
approaches of parameter estimation of the cosmological model will
work as well. Note that the highest certainty of parameter
estimation may be reached via a combination of independent tests
(for radio galaxies, see, e.g., \cite{vo_par1:Verkhodanov3_n_en,vo_par2:Verkhodanov3_n_en,vo_par3:Verkhodanov3_n_en}).

Describing the catalog in Papers I and II, we made a preliminary
statistical analysis, and constructed the Hubble diagrams,
describing the behavior of different parameters with the change in
redshift. In the present work, devoted to the construction of a
sample, and a statistical analysis of the radio galaxy catalog, we
collected the objects with known spectroscopic redshifts and with
measured angular sizes. To make the sample uniform, we adopted the
angular sizes from the NVSS catalog \cite{nvss:Verkhodanov3_n_en}, obtained on the
Very Large Array (VLA) with the beam of 45\arcsec.
In further studies we shall as well use the angular size data from
other surveys, like, for example, the FIRST survey \cite{first:Verkhodanov3_n_en}
data with the beam of 5\arcsec. Below we describe
our sample, make its statistical analysis and construct diagrams
for various physical parameters depending on redshifts.

\section{THE CATALOG}

\subsection{Description of the Catalog}

The data are presented in the Table, where the columns give the
object names, the equatorial coordinates of the galaxies from the
NVSS survey for epoch J2000.0, linear sizes (or upper limits) on
two axes in arc seconds, flux densities in mJy according to the
NVSS data, and redshifts. Sky positions of selected radio sources
are shown in Fig.\,1. The complete catalog is available from the
CATS database\footnote{\tt ftp://cats.sao.ru/pub/CATS/RGLIST},
and CDS ftp\footnote{\tt ftp://cdsarc.u-strasbg.fr/pub/cats/J/other/AstBu/64.357},
CDS database\footnote{\tt http://cdsarc.u-strasbg.fr/cgi-bin/VizieR?-source=J/other/AstBu/64.357}.
\begin{figure}[tbp]
\centerline{\hbox{
\psfig{figure=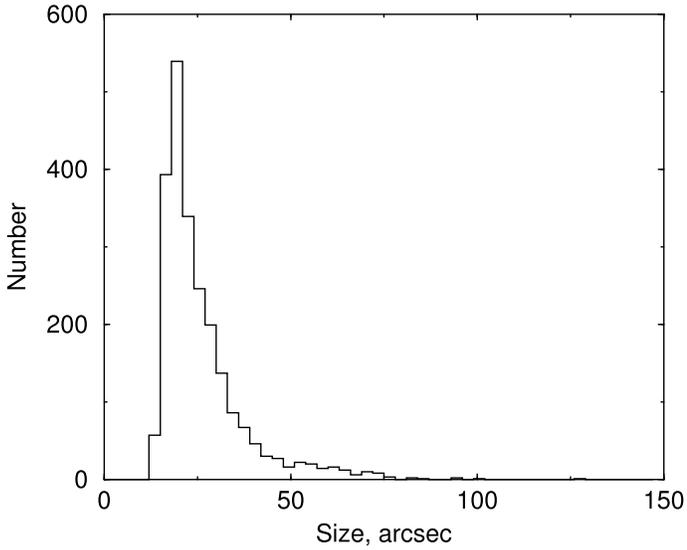,width=9cm,angle=-90}
}}
\caption{Histogram of the source angular sizes distribution.}
\label{f2:Verkhodanov3_n_en}
\end{figure}

\subsection{Statistical Analysis of the Sample}

We already pointed out in previous papers that our list of
galaxies is not complete and homogeneous, as the sample contains
objects drawn from different catalogs and unconnected sky areas.
The most complete and homogeneous subsample of the list contains
objects from the SDSS  survey, marked in Fig.~1 by white circles.
The SDSS radio galaxies include many objects with small redshifts
($z<0.5$) and small flux densities \mbox{($S<15$\,mJy)}, making
this subsample distinguish itself quality-wise among other
subsamples.

Using the resulting catalog, we made a preliminary statistical
analysis of the chosen objects. \mbox{Figs.\,2--4} demonstrate the
results of this analysis.

Figure\,2 represents the histogram of the sources major axes
distribution by angular size. The histogram peak's position de
facto corresponds to the size of point sources in the NVSS survey
after their reduction with the beam of 45\arcsec.

We constructed the ``source angular size--redshift--flux density''
(Fig.\,3) and ``angular size--redshift--spect- \linebreak ral
index'' (Fig.\,4) diagrams, and some projections onto the
parametric planes thereof. Spectral index was computed for the
frequency of 1400\,MHz in \linebreak Paper\,I.

For the ratio ``source size--redshift--flux density'', we made two
projections in the planes ``source size--flux density'', and
``source size--redshift''. For the first projection
``$\theta-S$'', we isolated three areas with different radio
source densities, corresponding to the unresolved sources from the
NVSS, extended radio objects, and gigantic radio galaxies. The
area boundaries can be provisionally described by the curves
(Fig.3, solid lines): (1) separating the unresolved and extended
radio galaxies as \mbox{$\theta = 28.3 \times \exp(-3p) + 19.3$,}
where $p = \ln(S/8.4)$, $\theta$ is the angular size in arc
seconds, $S$ is the flux density in mJy; (2) separating the
extended and gigantic radio galaxies as \mbox{$\theta = 97.4 - 5.5
\ln (S)$.} Different symbols mark different redshift ranges:
circles represent the $0.3<z<0.7$ range, squares mark the
\mbox{$0.7\le z<1.5$} range, pluses mark $1.5\le z$. Note that the
bulk of far galaxies from our sample (marked with bright gray) is
located at the flux density range from 40 to 300 mJy at 1400\,MHz.

The bottom source separation on the ``source size--flux density''
diagram is of course preconditioned by the selection effect,
defined by the size of the VLA directivity pattern in the NVSS
survey. For the study of sources from this area some better
resolution data are needed, for example, from the FIRST
\cite{first:Verkhodanov3_n_en} survey, as well conducted on the VLA, but with the
beam of 5'', or from the observations of different sources on VLBA
or VLBI. Besides, the NVSS data may contain some particularities,
linked with the source size determination at the noncircular
directivity pattern \cite{kimball:Verkhodanov3_n_en}. Nevertheless, the parameters
of the objects (extended radio sources) from the top and middle
parts of the diagram are applicable for further use in
cosmological tests.

On the second projection ``source size--redshift''
(``$\theta-z$''), we separated two object populations, marked with
different symbols: triangles mark the SDSS sources, and circles
indicate the rest of the objects. One can see from the figure that
the SDSS radio galaxies mainly belong to the small $z$ range in
our sample, having a flux density scatter from the minimal to the
maximal values. Note that the bigger radio galaxies are of
interest in estimating the extended objects contribution into the
microwave \mbox{background \cite{grg_piter08:Verkhodanov3_n_en}.}

\begin{figure*}[tbp]
\centerline{\vbox{
\hbox{
\mbox{\hspace*{3cm}}
\psfig{figure=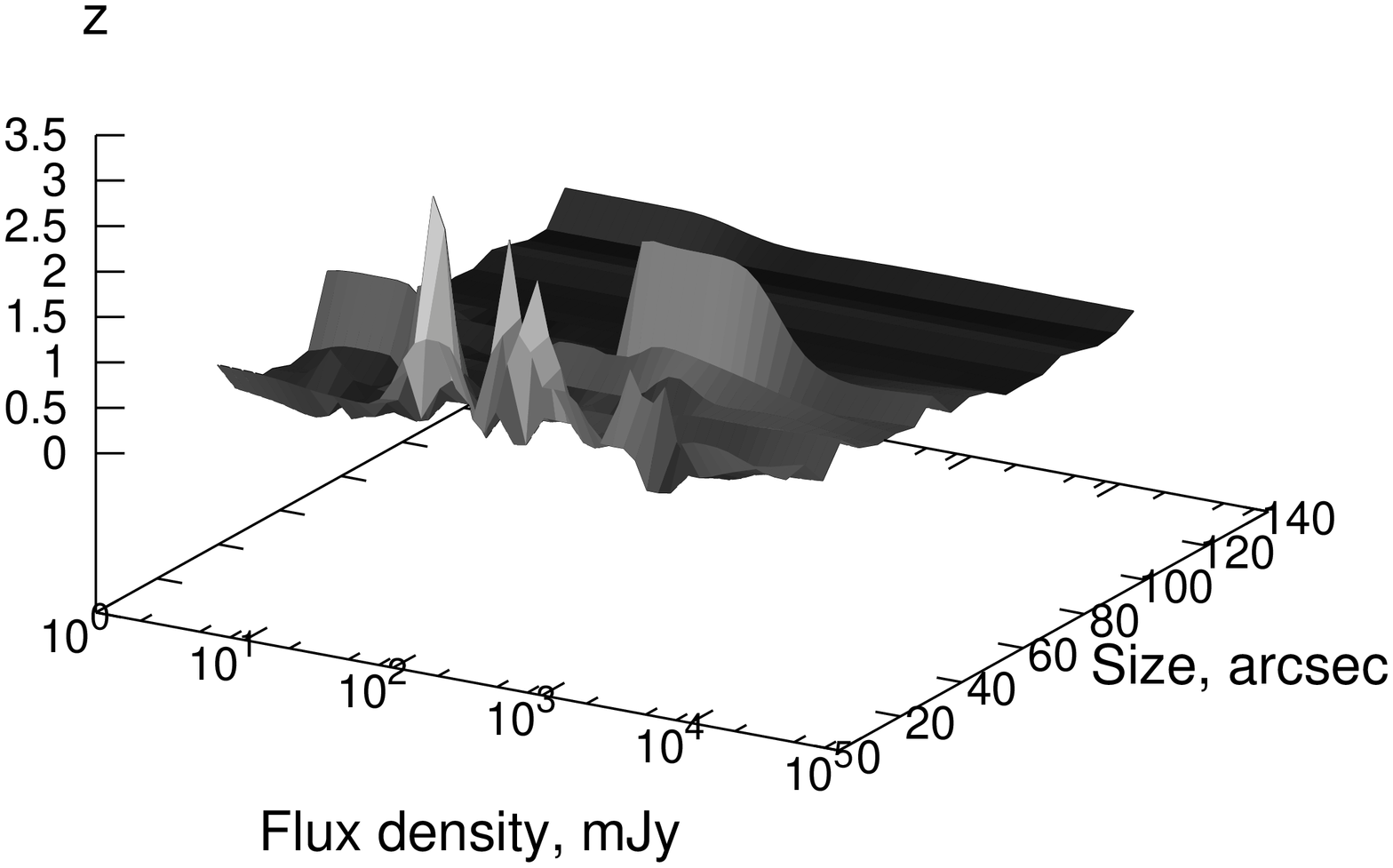,width=9cm,angle=-90}
}
\hbox{
\psfig{figure=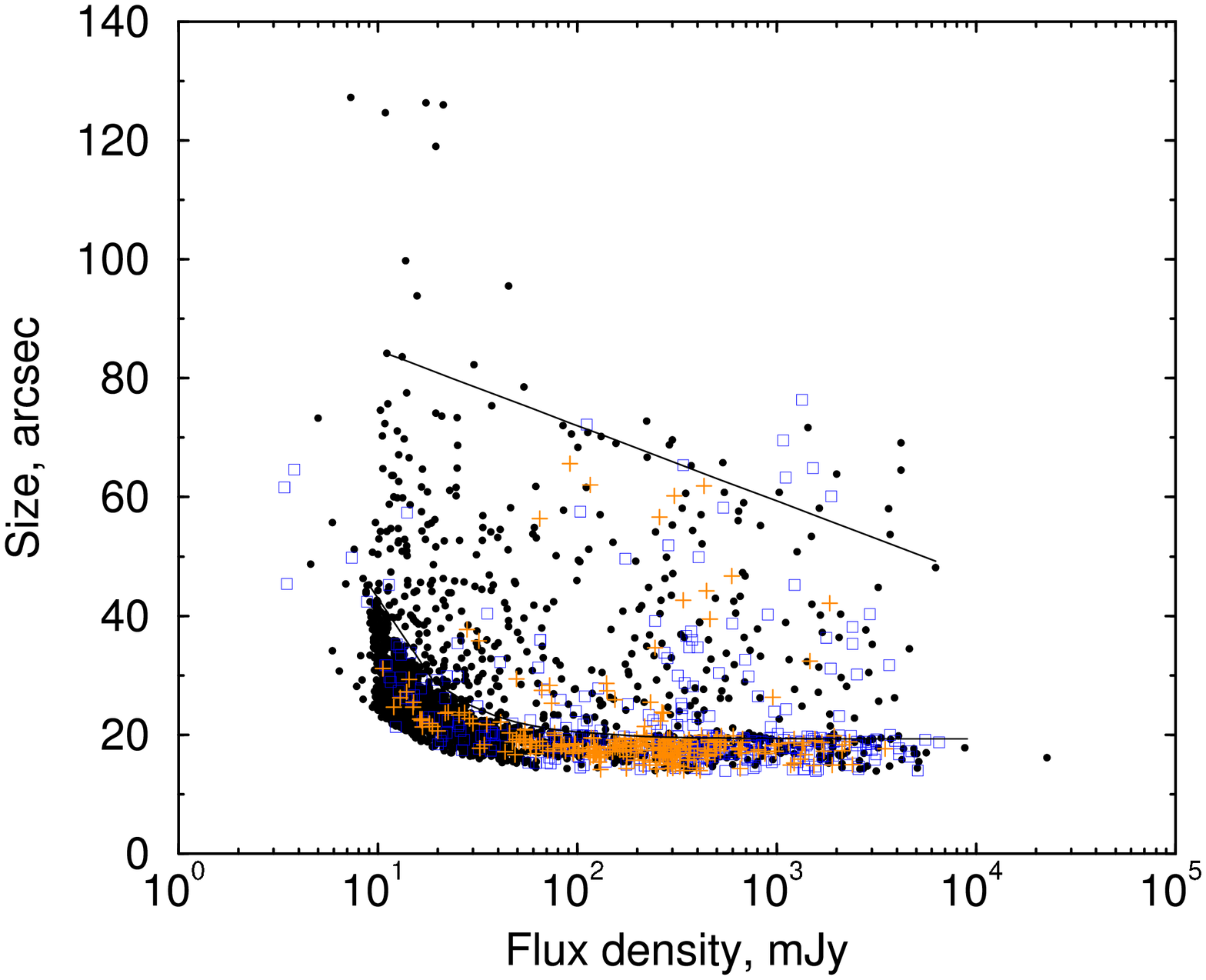,width=8cm,angle=-90}
\psfig{figure=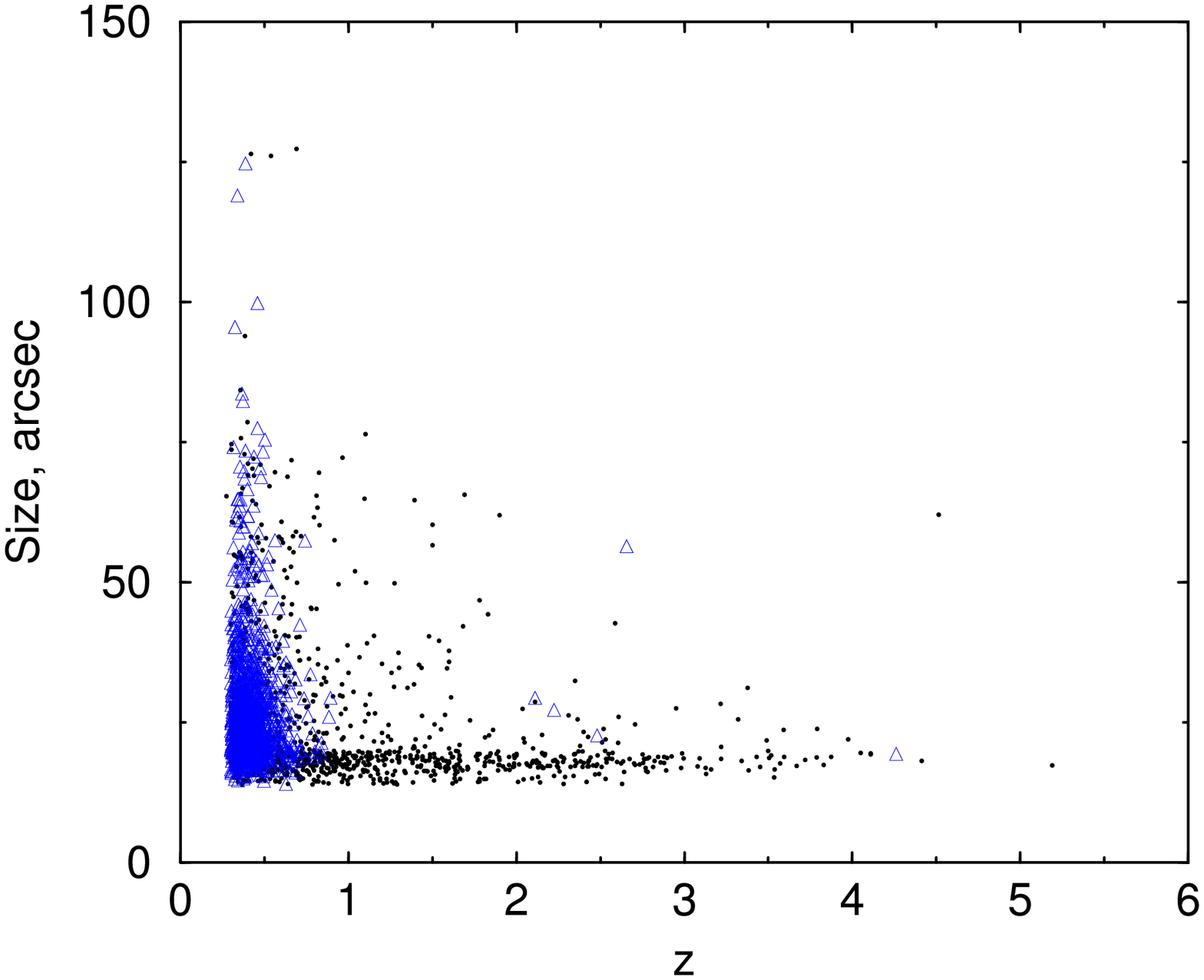,width=7.7cm,angle=-90}
} }}
\caption{Top: ``angular size--redshift--flux density''
relation for radio galaxies. Bottom left: ``source size--flux
density'' relation. Different symbols mark different redshift
ranges: circles---$0.3<z<0.7$, squares---$0.7\le z<1.5$,
pluses---$1.5\le z$. Solid lines divide the areas of different
density. Bottom right: ``source size--redshift'' relation.
Triangles indicate the SDSS data. Circles mark the objects from
all other catalogs.}
\label{f3:Verkhodanov3_n_en}
\end{figure*}

% Fig.4
\begin{figure*}[tbp]
\centerline{\vbox{
\hbox{
\mbox{\hspace*{3cm}}
\psfig{figure=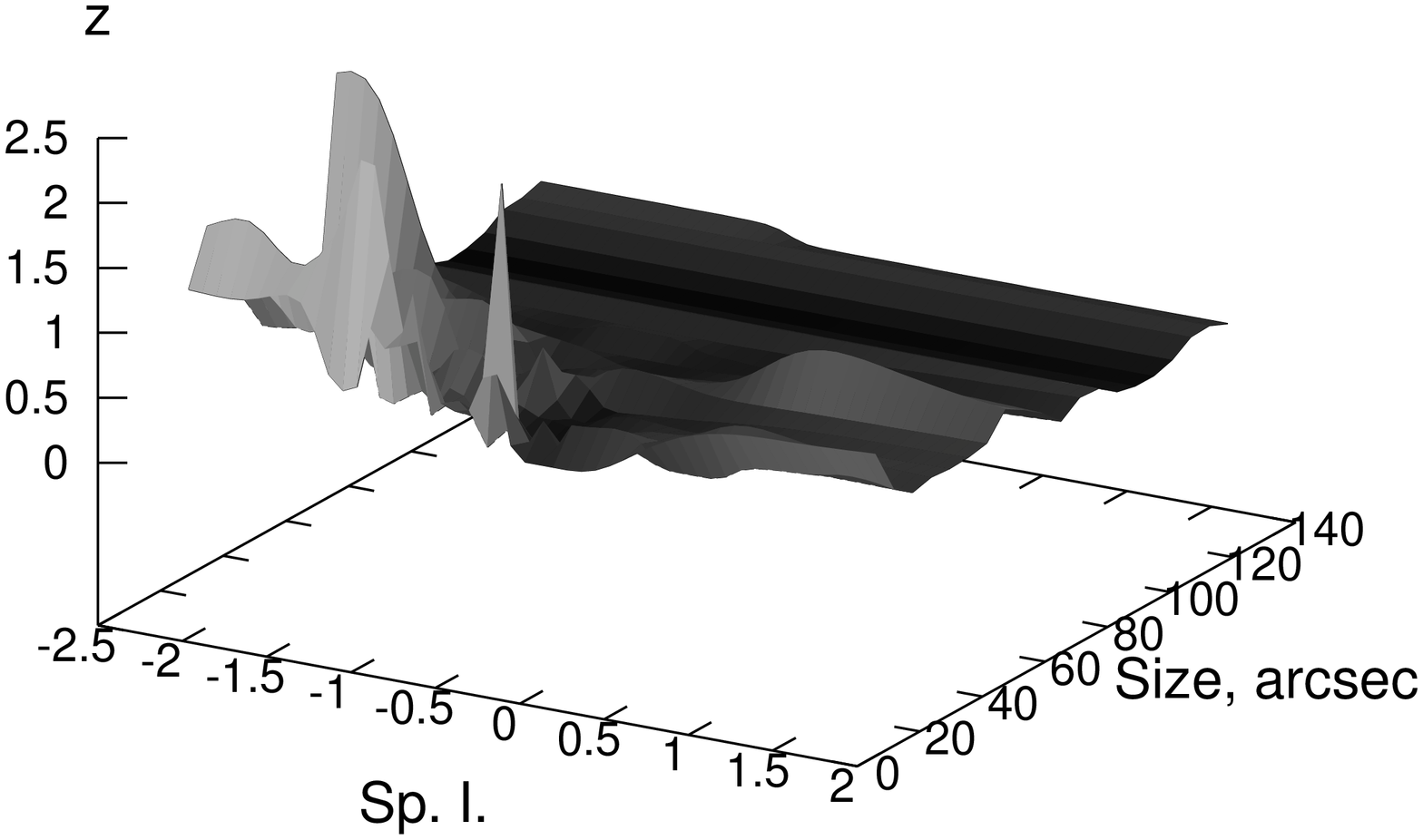,width=9cm,angle=-90}
}
\hbox{
\psfig{figure=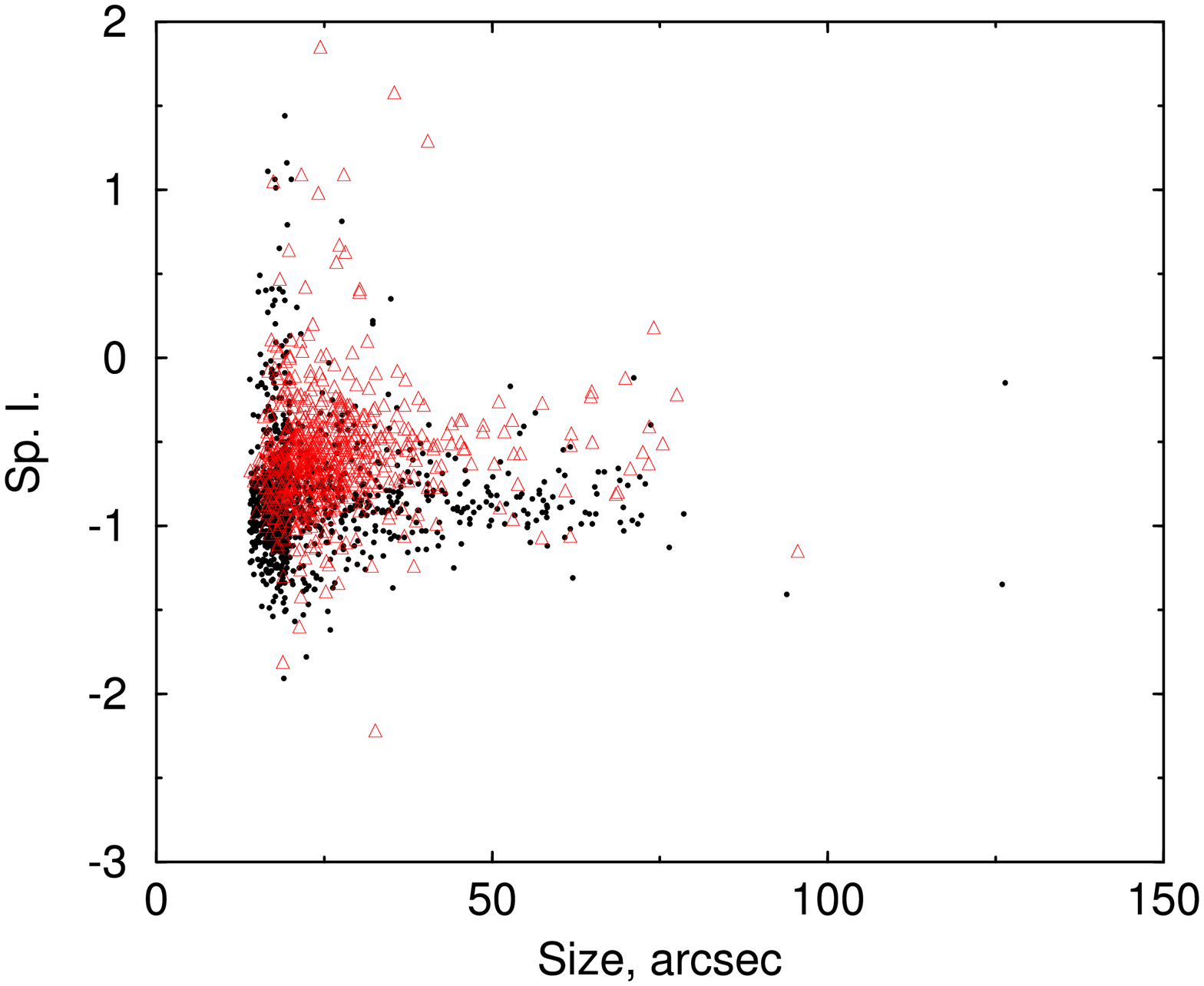,width=8cm,angle=-90}
\psfig{figure=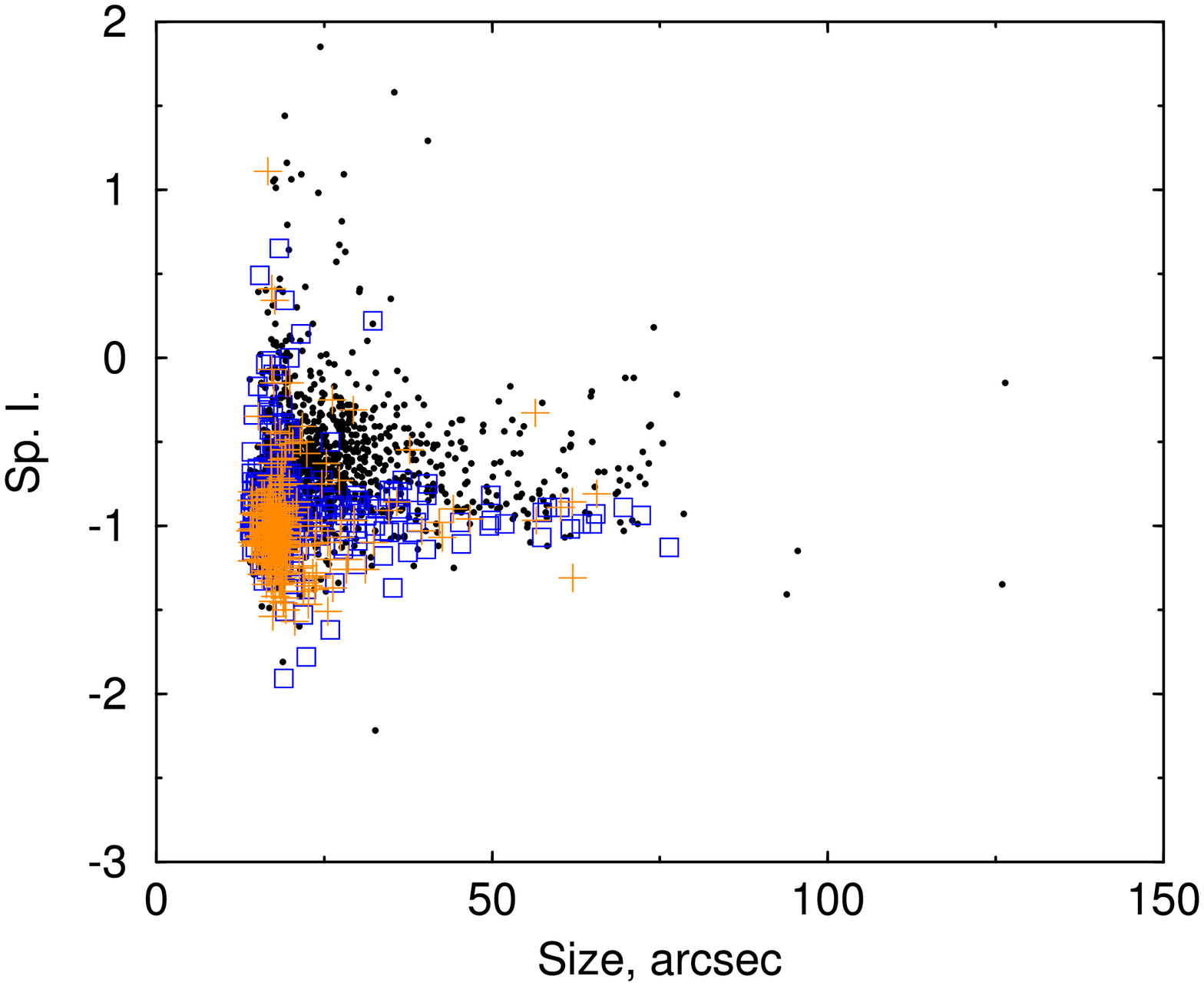,width=8cm,angle=-90}
} }}
\caption{Top: ``source size--redshift--spectral index''
relation for radio galaxies. The spectral index is computed at
1400\,MHz. Bottom left: ``spectral index--source size'' relation.
Triangles mark the SDSS data, circles represent the objects from
all other catalogs. Bottom right: again, ``spectral index--source
size'' relation. Different symbols mark different redshift ranges:
circles---$0.3<z<0.7$, squares---$0.7\le z<1.5$, pluses---$1.5\le
z$. }
\label{f4:Verkhodanov3_n_en}
\end{figure*}

For the ``source size--redshift--spectral index'' relation
(Fig.\,4) we made only one projection---``spectral index--source
size'', presented in two variants. In the first one the triangles
mark the SDSS objects, in the second one different symbols mark
\linebreak objects with different redshift ranges: \linebreak
circles---\mbox{$0.3<z<0.7$,} squares---$0.7\le z<1.5$, \linebreak
pluses mark the objects with $1.5\le z$.  The diagram demonstrates
a relationship between the redshift of the source and the spectral
index in radio range, discovered long ago
\cite{dagkes:Verkhodanov3_n_en,blum_miley:Verkhodanov3_n_en}, what was already discussed in Paper~I.
For another thing, we observe the expected relation---bigger
sources have steeper spectra due to the synchrotron emission in
radio lobes. Note that for the radio galaxies discovered in the
SDSS, the value of the spectral index is smaller, while the
distribution on the source size does not bear any great
differences. This difference is explained by the selection effect
and an existence of the `spectral index--redshift'' relation. This
latter fact is demonstrated on the right bottom figure from
Fig.\,4, where the region of large $z$ is marked in light gray.

\section{CONCLUSIONS}

We report the final part of the radio galaxies catalog with
redshifts $z>0.3$, which contains complete information on the
angular sizes of the objects, and flux densities at 1400\,MHz. The
catalog is based on lists of objects from the NED and CATS
databases. Our complete sample includes the SDSS objects, and the
objects from the ``Big Trio'' program \cite{par_big3a:Verkhodanov3_n_en,par_big3b:Verkhodanov3_n_en}.
We perform a preliminary statistical analysis of the angular size
data, flux density and spectral index calculations. On the diagram
``source size--flux density'' we separated three radio galaxy
subsamples, varying in angular sizes and flux densities. We
approximate the boundaries of these subsamples based on analytic
functions. The bottom object separation on the ``source
size--flux density'' diagram is preconditioned by the selection
effect defined by the size of the VLA directivity pattern in the
NVSS survey. In order to use the sources from this area in the
cosmological tests, connected with the source size, some better
resolution data are needed.

%\begin{acknowledgments}
\noindent
{\small
{\bf Acknowledgments}.

This research has made use of the NASA/IPAC Extragalactic Database
(NED) which is operated by the Jet Propulsion Laboratory,
California Institute of Technology, under contract with the
National Aeronautics and Space Administration. We also made use of
the CATS database of radio astronomical catalogs \cite{cats:Verkhodanov3_n_en,cats2:Verkhodanov3_n_en}
and  FADPS\footnote{\tt http://sed.sao.ru/$\sim$vo/fadps\_e.html}
radio-astronomical data reduction  system~\cite{fadps:Verkhodanov3_n_en,fadps2:Verkhodanov3_n_en}. The
authors are grateful to Olga Kashibadze (the SAO RAS) for the aid
in nonlinear function approximation using the MATLAB package. This
work was supported by the Program of the Support of Leading
Scientific Schools in Russia (the School of S.\,M.\,Khaikin) and
the grants of the Russian Foundation of Basic Research (project
nos. 09-02-00298 and 09-02-92659-IND). O.V.V. acknowledges partial
support from the RFBR (project \mbox{No\,07-02-01417-a}), and the
Fund for Support of National Science (``Young PhDs of the RAS''
program).
}

\end{document}